# Degenerate solutions to the Dirac and Weyl equations and their applications

Georgios N. Tsigaridas[1,*], Aristides I. Kechriniotis[2], Christos A. Tsonos[2] and Konstantinos K. Delibasis[3]

[1]Department of Physics, School of Applied Mathematical and Physical Sciences, National Technical University of Athens, GR-15772 Zografou Athens, Greece

[2]Department of Physics, University of Thessaly, GR-35100 Lamia, Greece

[3]Department of Computer Science and Biomedical Informatics, University of Thessaly, GR-35131 Lamia, Greece

[*]Corresponding Author. E-mail: gtsig@mail.ntua.gr

**Abstract**

In this review article we present a comprehensive review of degenerate solutions to the Dirac and Weyl equations, highlighting novel and significant findings. Specifically, we demonstrate that all Weyl particles, and under certain conditions Dirac particles, can occupy the same quantum state under an extensive range of electromagnetic 4-potentials and fields. These fields, which are infinite in number, have been explicitly derived and analysed. Additionally, we establish that Weyl particles can form localized states even in the absence of external electromagnetic fields. Moreover, we show that their localization can be precisely controlled through the application of simple electric fields, offering a tuneable mechanism for manipulating these particles. Building on these insights, we propose an innovative device that utilizes Weyl fermions to control the flow of information at an unprecedented rate of up to 100 petabits per second. This finding has significant implications regarding the development of next-generation electronic and quantum information technologies, as it presents a fundamentally new approach to high-speed data processing and transmission.

**Keywords**: Dirac equation, Weyl equation, Degenerate solutions, Dirac particles, Weyl particles, Electromagnetic 4-potentials, Electromagnetic fields, Electromagnetic waves, Information control, Quantum tunnelling, Quantum computing

## 1. Introduction

Let us consider Dirac equation in the following form

$$i\gamma^{\mu}\partial_{\mu}\Psi + a_{\mu}\gamma^{\mu}\Psi - m\Psi = 0 \tag{1.1}$$

where $\gamma^{\mu}$ are the four contravariant gamma matrices in Dirac representation

$$\gamma^0 = \begin{pmatrix} \sigma^0 & 0 \\ 0 & -\sigma^0 \end{pmatrix} \qquad \gamma^\mu = \begin{pmatrix} 0 & \sigma^\mu \\ -\sigma^\mu & 0 \end{pmatrix} \tag{1.2}$$

and $m$ is the mass of the particle. Here $\sigma^\mu$ are the well-known Pauli matrices in the following form

$$\sigma^0 = \begin{pmatrix} 1 & 0 \\ 0 & 1 \end{pmatrix} \quad \sigma^1 = \begin{pmatrix} 0 & 1 \\ 1 & 0 \end{pmatrix} \quad \sigma^2 = \begin{pmatrix} 0 & -i \\ i & 0 \end{pmatrix} \quad \sigma^3 = \begin{pmatrix} 1 & 0 \\ 0 & -1 \end{pmatrix} \tag{1.3}$$

and $a_\mu = qA_\mu$, where $q$ is the charge of the particle and $A_\mu$ is the electromagnetic 4-potential. It should also be noted that Eq. (1.1) is expressed in natural units, where $\hbar = c = 1$.

An intriguing question raised by Dirac to Eliezer [1] in 1958 concerns the relationship between a given wave function and the corresponding electromagnetic 4-potential as dictated by Dirac's equation. Specifically, given a known wave function, to what extent is the associated 4-potential uniquely determined? If it is not uniquely specified, what degree of arbitrariness exists in its definition?

This question was partially answered by Booth, Legg and Jarvis [2] in 2001, showing that if $\Psi^\dagger \gamma \, \Psi \neq 0$, then $\Psi$ corresponds to a unique 4-potential defined by the formula

$$a_\mu = \frac{i}{2} \frac{\bar{\Psi}\gamma^5\gamma^\mu \not{\partial}\Psi - \bar{\Psi}\gamma^5 \overleftarrow{\not{\partial}}\gamma^\mu\Psi}{\bar{\Psi}\gamma^5\Psi} \tag{1.4}$$

where

$$\gamma^5 = i\gamma^0\gamma^1\gamma^2\gamma^3, \; \gamma = \gamma^0 + \gamma^0\gamma^5, \; \bar{\Psi} = \Psi^\dagger\gamma^0, \; \not{\partial} = \gamma^\mu\partial_\mu, \; \bar{\Psi}\gamma^\nu \overleftarrow{\not{\partial}}\gamma^\mu\Psi = \overline{\not{\partial}\Psi}\,\gamma^\nu\gamma^\mu\Psi \tag{1.5}$$

However, the question what happens if $\Psi^\dagger \gamma \, \Psi = 0$ remained open until 2020, where we have shown that, in this case, the wavefunction $\Psi$ is degenerate, corresponding to an infinite number of electromagnetic 4-potentials which are explicitly calculated in Theorem 5.4 in [3]. Specifically, they are given by the formula

$$b_\mu = a_\mu + s(\mathbf{r}, t)\theta_\mu \tag{1.6}$$

where $s(\mathbf{r}, t)$ is any real function of the spatial coordinates and time and

$$(\theta_0, \theta_1, \theta_2, \theta_3) = \left(1, -\frac{\Psi^T\gamma^0\gamma^1\gamma^2\Psi}{\Psi^T\gamma^2\Psi}, -\frac{\Psi^T\gamma^0\Psi}{\Psi^T\gamma^2\Psi}, \frac{\Psi^T\gamma^0\gamma^2\gamma^3\Psi}{\Psi^T\gamma^2\Psi}\right) \tag{1.7}$$

It should be noted that the above formula is valid under the condition that $\Psi^T\gamma^2\Psi \neq 0$. In the case that $\Psi^T\gamma^2\Psi = 0$ and $\Psi^\dagger \gamma \, \Psi = 0$, it has been proven [3] that the wavefunction $\Psi$ can be written either in the form

$$\Psi = \begin{pmatrix} \psi \\ \psi \end{pmatrix} \tag{1.8}$$

where $\psi$ is solution to the Weyl equation

$$a_\mu \sigma^\mu \psi = -i\sigma^\mu \partial_\mu \psi \tag{1.9}$$

describing massless particles with spin parallel to their propagation direction (positive helicity), or in the form

$$\Psi = \begin{pmatrix} \psi \\ -\psi \end{pmatrix} \tag{1.10}$$

where $\psi$ is solution to the Weyl equation

$$2a_0\sigma^0\psi - a_\mu \sigma^\mu \psi = -\left(2i\sigma^0\partial_0\psi - i\sigma^\mu \partial_\mu \psi\right) \tag{1.11}$$

describing massless particles with spin anti-parallel to their propagation direction (negative helicity). Furthermore, as shown in Theorem 3.1 in [3], all the solutions to the Weyl equations are degenerate, corresponding to an infinite number of real 4-potentials, given by the formulae

$$b_\mu = a_\mu + s(\mathbf{r},t)\varphi_\mu \tag{1.12}$$

where

$$(\varphi_0,\varphi_1,\varphi_2,\varphi_3) = \left(1, \pm\frac{\psi^\dagger\sigma^1\psi}{\psi^\dagger\psi}, \pm\frac{\psi^\dagger\sigma^2\psi}{\psi^\dagger\psi}, \pm\frac{\psi^\dagger\sigma^3\psi}{\psi^\dagger\psi}\right) \tag{1.13}$$

Here, the plus and minus sign corresponds to the cases of negative and positive helicity respectively.

In the following, we shall discuss some degenerate solutions for massless Dirac and Weyl particles (section 2), some degenerate solutions for massive particles associated with quantum tunnelling (section 3), some degenerate wavelike solutions for massive particles (section 4) and a general method for obtaining degenerate solutions for massive, massless and Weyl particles (section 5). In section 6 we shall also discuss some interesting solutions to the Weyl equations regarding the existence of localized states in the absence of electromagnetic fields and the use of simple electric fields for controlling the localization of Weyl particles. Based on these results, in section 7, we propose a device for controlling the flow of information with rates up to 100 Petabits per second, based on Weyl fermions. Finally, our conclusions are presented in section 8.

## 2. Degenerate solutions for massless Dirac and Weyl particles

In [3] it has been shown that that the wavefunction $\Psi$ of a free Dirac particle is degenerate if and only if the particle is massless. In this case, $\Psi$ corresponds to an infinite number of real 4-potentials of the form

$$(a_0, a_1, a_2, a_3) = (1, -\sin\theta\cos\varphi, -\sin\theta\sin\varphi, -\cos\theta)\, s(\mathbf{r}, t) \quad (2.1)$$

where $s(\mathbf{r}, t)$ is an arbitrary real function of the spatial coordinates and time and $(\theta,\ \varphi)$ are the angles defining the propagation direction of the particles in spherical coordinates. The electromagnetic fields corresponding to the above 4-potentials can be easily calculated through the formula [4]

$$\mathbf{E} = -\nabla U - \frac{\partial \mathbf{A}}{\partial t}, \qquad \mathbf{B} = \nabla \times \mathbf{A} \quad (2.2)$$

where $U = a_0/q$ is the electric potential and $\mathbf{A} = -(1/q)(a_1\mathbf{i} + a_2\mathbf{j} + a_3\mathbf{k})$ is the magnetic vector potential. The choice of the minus sign in the definition of magnetic potential is related to the form of the Dirac equation used in this article.

Indeed, using Eq. (2.2) we obtain that

$$\mathbf{E}(\mathbf{r}, t) = -\nabla s_q - \frac{\partial s_q}{\partial t}(\sin\theta\cos\varphi\ \mathbf{i} + \sin\theta\sin\varphi\ \mathbf{j} + \cos\theta\ \mathbf{k}) \quad (2.3)$$

and

$$\begin{aligned}\mathbf{B}(\mathbf{r}, t) = &\left(\cos\theta\frac{\partial s_q}{\partial y} - \sin\theta\sin\varphi\frac{\partial s_q}{\partial z}\right)\mathbf{i} - \left(\cos\theta\frac{\partial s_q}{\partial x} - \sin\theta\cos\varphi\frac{\partial s_q}{\partial z}\right)\mathbf{j} + \\ &\sin\theta\left(\sin\varphi\frac{\partial s_q}{\partial x} - \cos\varphi\frac{\partial s_q}{\partial y}\right)\mathbf{k}\end{aligned} \quad (2.4)$$

where $s_q = s(\mathbf{r}, t)/q$. Some special cases of practical interest are the following.

If we suppose that the arbitrary function $s_q$ depends only on time, then $\mathbf{B}(\mathbf{r}, t) = \mathbf{0}$ and the electric field is simplified, taking the form

$$\mathbf{E}(\mathbf{r}, t) = -\frac{ds_q}{dt}(\sin\theta\cos\varphi\mathbf{i} + \sin\theta\sin\varphi\mathbf{j} + \cos\theta\mathbf{k}) \quad (2.5)$$

The same outcome holds for free Weyl particles. As a result, the state of these particles remains unaffected by an electric field of any strength or time dependence when applied along their direction of motion. This implies that the electric current carried by charged particles in degenerate states remains unchanged, even when a voltage of any magnitude or time dependence is applied in the same direction, in contrast to what is predicted by Ohm's law in classical physics. However, for charged particles in non-degenerate states, classical physics is valid, and Ohm's law dictates that the electric current is proportional to the applied voltage. Thus, the relationship between electric current and applied voltage differs fundamentally between degenerate and non-degenerate states. This distinction provides a means to experimentally identify degenerate states in materials hosting massless Dirac or Weyl particles, such as graphene sheets [5-11] and Weyl semimetals [12–31].

On the other hand, if the arbitrary function $s_q$ is of the form

$$s_q(\mathbf{r},t) = -E_{w1}\cos\left[k_w(z-t)+\delta_{w1}\right]x - E_{w2}\cos\left[k_w(z-t)+\delta_{w2}\right]y \tag{2.6}$$

the corresponding electromagnetic field becomes

$$\mathbf{E}(\mathbf{r},t) = E_{w1}\cos\left[k_w(z-t)+\delta_{w1}\right]\mathbf{i} + E_{w2}\cos\left[k_w(z-t)+\delta_{w2}\right]\mathbf{j} \tag{2.7}$$

$$\mathbf{B}(\mathbf{r},t) = -E_{w2}\cos\left[k_w(z-t)+\delta_{w2}\right]\mathbf{i} + E_{w1}\cos\left[k_w(z-t)+\delta_{w1}\right]\mathbf{j} \tag{2.8}$$

describing a plane electromagnetic wave propagating along the same direction with the particles. Here, $E_{w1},\delta_{w1},E_{w2},\delta_{w2}$ are arbitrary real constants corresponding to the amplitude and phase of the electromagnetic wave and $k_w$ is a real parameter corresponding to the wavenumber.

As a result, the state of free massless Dirac or Weyl particles remains unaffected by plane electromagnetic waves, such as laser beams, regardless of their polarization, when propagating along the particles' direction of motion. This means that particles in degenerate states and electromagnetic waves can travel in the same direction without interacting, unlike charged particles in non-degenerate states, which do interact with the electromagnetic field.

This property can be utilized to detect the presence of degenerate states using a Michelson interferometer [32]. Specifically, if a material containing charged particles in degenerate states is placed in one arm of the interferometer, the electromagnetic wave traveling through it will behave as if it were moving through a vacuum. Conversely, if the particles are in non-degenerate states, they will interact with the wave, altering its velocity and thus its phase. Consequently, the transition between degenerate and non-degenerate states can be readily observed through changes in the interference pattern generated by the interferometer, as illustrated in Figure 1.

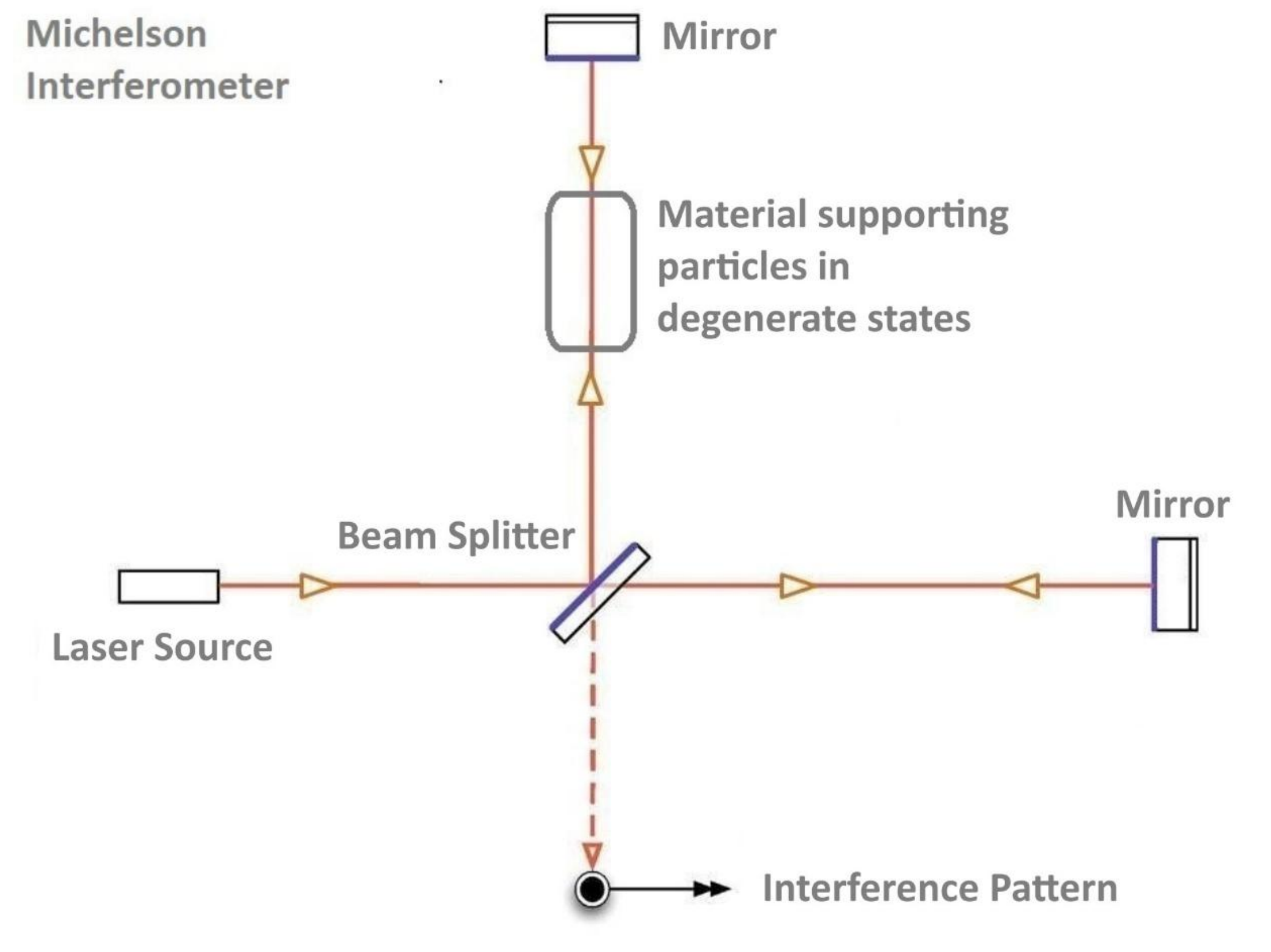

**Figure 1**: A proposed method for experimentally detecting the presence of degenerate states using a Michelson interferometer [32].

It is important to note that not all particles will travel exactly parallel to the electromagnetic wave. As a result, some will still interact with the wave, even in degenerate states. However, interferometric techniques are highly sensitive, meaning that even a slight reduction in particle-wave interaction should be enough to produce a detectable shift in the interference pattern, as shown in [32].

Additionally, recent work by Shao et al. [33] has confirmed the existence of semi-Dirac fermions—particles that exhibit mass only when moving in a specific direction. These fparticles are expected to remain in degenerate states exclusively when traveling along the direction where their effective mass is zero. Consequently, if a material hosting semi-Dirac fermions is placed within the Michelson interferometer depicted in Figure 1, the interference pattern should vary depending on the material's orientation. This implies that by simply rotating the material inside the interferometer, one can observe the transition between degenerate and non-degenerate states as the particles gain mass.

Furthermore, in our effort to find more general forms of degenerate solutions to the Dirac equation [34], we have shown that all spinors of the form

$$\Psi_p = \left(c_1 u_\uparrow + c_2 u_\downarrow\right)\exp\left[ih\left(x,y,z,t\right)\right],\ \Psi_a = \left(c_1 v_\uparrow + c_2 v_\downarrow\right)\exp\left[ih\left(x,y,z,t\right)\right] \quad (2.9)$$

are degenerate, corresponding to massless Dirac particles or antiparticles propagating along a direction in space defined by the angles $(\theta,\ \varphi)$ in spherical coordinates. Here, $c_1, c_2$ are arbitrary complex constants, $h(x,y,z,t)$ is any real function of the spatial coordinates and time, and $u_\uparrow, u_\downarrow, v_\uparrow, v_\downarrow$ are eigenvectors describing the spin state of the particle $(u_\uparrow, u_\downarrow)$ or antiparticle $(v_\uparrow, v_\downarrow)$ [35].

The 4-potential corresponding to the above solutions is given by the following simple formula [34]

$$(a_0, a_1, a_2, a_3) = \left( \frac{\partial h}{\partial t}, \frac{\partial h}{\partial x}, \frac{\partial h}{\partial y}, \frac{\partial h}{\partial z} \right) \tag{2.10}$$

In addition, according to Theorem 5.4 in [3], spinors given by Eq. (2.9) will also be solutions to the Dirac equation for an infinite number of 4-potentials, given by the formula

$$b_\mu = a_\mu + s(\mathbf{r},t)\kappa_\mu \tag{2,11}$$

where

$$\begin{aligned}(\kappa_0,\kappa_1,\kappa_2,\kappa_3) &= \left(1, -\frac{\Psi^T\gamma^0\gamma^1\gamma^2\Psi}{\Psi^T\gamma^2\Psi}, -\frac{\Psi^T\gamma^0\Psi}{\Psi^T\gamma^2\Psi}, \frac{\Psi^T\gamma^0\gamma^2\gamma^3\Psi}{\Psi^T\gamma^2\Psi}\right) \\ &= (1, -\sin\theta\cos\varphi, -\sin\theta\sin\varphi, -\cos\theta)\end{aligned} \tag{2.12}$$

In Gaussian units, the electromagnetic fields associated with the aforementioned 4-potentials are identical to those corresponding to free Dirac particles, as expressed by equations (2.3) and (2.4). Therefore, as in the case of free Dirac particles, the state of the particles described by the more general solutions given by Eq. (2.9) remains unaffected by an electric field with arbitrary time-dependence, when applied along the direction of motion of the particles. Likewise, particles in the degenerate states described above will not interact with a plane electromagnetic wave, such as a laser beam, regardless of its polarization, if it propagates parallel to their direction of motion.

In addition, an intriguing aspect of degeneracy, as defined in [3], is the transition from degenerate to non-degenerate states as particles gain mass. In [34], it has been shown that, for massive particles, the spinors in Eq. (2.9) remain solutions to the Dirac equation under the 4-potentials in Eq. (2.10) but no longer satisfy the massless Dirac equation for the more general 4-potentials in Eq. (2.11). Indeed, substituting these spinors into the massless Dirac equation and using the 4-potentials given by Eq. (2.11), we obtain that

$$i\gamma^\mu\partial_\mu\Psi_p + b_\mu\gamma^\mu\Psi_p = \left(1 - \sqrt{\frac{1-e}{1+e}}\right) s\Psi_p^{\,*} \tag{2.13}$$

$$i\gamma^{\mu}\partial_{\mu}\Psi_{a}+b_{\mu}\gamma^{\mu}\Psi_{a}=-\left(1-\sqrt{\frac{1-e}{1+e}}\right)s\Psi_{a}^{*} \tag{2.14}$$

where $\Psi_{p}^{*}$, $\Psi_{a}^{*}$ are the unperturbed spinors for particles and antiparticles respectively and $e$ is the ratio of the rest energy of the particles to their total energy. Thus, in the case that the rest energy of the particles is much smaller than their total energy $(e<<1)$, the above equations take the simpler form

$$i\gamma^{\mu}\partial_{\mu}\Psi_{p}+b_{\mu}\gamma^{\mu}\Psi_{p}=e\,s\Psi_{p}^{*} \tag{2.15}$$

$$i\gamma^{\mu}\partial_{\mu}\Psi_{a}+b_{\mu}\gamma^{\mu}\Psi_{a}=-e\,s\Psi_{a}^{*} \tag{2.16}$$

This result implies that as the ratio of the rest energy to the total energy of the particles increases, the function $s$ should be restricted to smaller values, suppressing the effects of degeneracy. On the other hand, as the ratio $e$ decreases, the function $s$ is allowed to take larger values, and the degeneracy becomes more evident. Finally, as $e$ tends to zero, there is no restriction on the values of the function $s$ and the theory of degeneracy becomes fully applicable.

Finally, it should be mentioned that the ratio $e$ becomes also negligible if the total energy of the particles is much higher than their rest energy. Thus, the theory of degeneracy is also expected to be valid, in an approximate sense, for high energy particles. Furthermore, the higher the total energy of the particles compared to their rest energy, the more evident the effects of degeneracy are expected to become. More details on this very interesting remark can be found in the Appendix of [36].

## 3. Degenerate solutions to the Dirac equation for massive particles and their applications in quantum tunnelling

All the previous results consider degenerate solutions for massless Dirac and Weyl particles. However, an interesting question is the following: Are there degenerate solutions for massive particles, and if yes, what is their physical interpretation?

Indeed, it has been shown that all spinors of the form [37]

$$\Psi=c_{1}\exp\left(if\cos\xi\right)\exp\left[-\frac{m}{\sin^{2}\xi}\left(-z+t\cos\xi\right)\right]\begin{pmatrix} i\sin\xi \\ -i-\cos\xi \\ \sin\xi \\ 1+i\cos\xi \end{pmatrix} \tag{3.1}$$

are degenerate and satisfy the Dirac equation for the 4-potentials

$$\left(a_0, a_1, a_2, a_3\right) = \cos\xi\left(\frac{\partial f}{\partial t}, \frac{\partial f}{\partial x} - \frac{m}{\sin\xi}, \frac{\partial f}{\partial y}, \frac{\partial f}{\partial z}\right) \tag{3.2}$$

Here, $c_1$ is an arbitrary complex constant, $\xi$ a real parameter $\left(\xi \neq n\pi, n \in \mathbb{Z}\right)$ and $f$ any real function of the spatial coordinates and time. Furthermore, according to Theorem 5.4 in [3], the spinors (3.1) will also be solutions to the Dirac equation for an infinite number of 4-potentials, given by the formula

$$b_\mu = a_\mu + s\left(\mathbf{r}, t\right)\kappa_\mu \tag{3.3}$$

where

$$\left(\kappa_0, \kappa_1, \kappa_2, \kappa_3\right) = \left(1, -\frac{\Psi^T\gamma^0\gamma^1\gamma^2\Psi}{\Psi^T\gamma^2\Psi}, -\frac{\Psi^T\gamma^0\Psi}{\Psi^T\gamma^2\Psi}, \frac{\Psi^T\gamma^0\gamma^2\gamma^3\Psi}{\Psi^T\gamma^2\Psi}\right) = \left(1, 0, \sin\xi, -\cos\xi\right) \tag{3.4}$$

Setting $\xi = n\pi + \pi/2, n \in \mathbb{Z}$ the degenerate spinors given by Eq. (3.1) become time-independent, taking the simple form

$$\Psi = c_1 \exp\left(m\, z\right)\begin{pmatrix} \pm i \\ -i \\ \pm 1 \\ 1 \end{pmatrix} \tag{3.5}$$

which are also solutions to the one-dimensional time-independent Dirac equation

$$i\gamma^3\partial_z\Psi - m\Psi = 0 \tag{3.6}$$

for zero 4-potential. In general, it can be shown that all spinors of the following form [37]

$$\Psi_p = c_+ \exp\left(-m\, z\right)\begin{pmatrix} 1 \\ 1 \\ i \\ -i \end{pmatrix} + c_- \exp\left(m\, z\right)\begin{pmatrix} -1 \\ 1 \\ i \\ i \end{pmatrix} \tag{3.7}$$

$$\Psi_a = c_+ \exp\left(m\, z\right)\begin{pmatrix} i \\ -i \\ 1 \\ 1 \end{pmatrix} + c_- \exp\left(-m\, z\right)\begin{pmatrix} i \\ i \\ -1 \\ 1 \end{pmatrix} \tag{3.8}$$

are degenerate solutions to the Dirac equation for zero 4-potential. Here $c_+$, $c_-$ are arbitrary complex constants corresponding to motion along the +z and -z direction respectively.

These solutions can be interpreted in the framework of quantum tunnelling, as particles (Eq. 3.7) or antiparticles (Eq. 3.8) moving along the $\pm z$ direction in a potential barrier with height equal to the energy of the particles (or antiparticles). Also, the spin of the particles (or antiparticles) should be perpendicular to their direction of motion, to ensure equal contribution of the positive and negative helicity eigenstates. This condition is also satisfied for pairs of particles (Eq 3.7) or antiparticles (Eq. 3.8) with spin opposite to each other. A schematic representation of quantum tunneling for Dirac fermions in degenerate states is shown in Figure 2 [37].

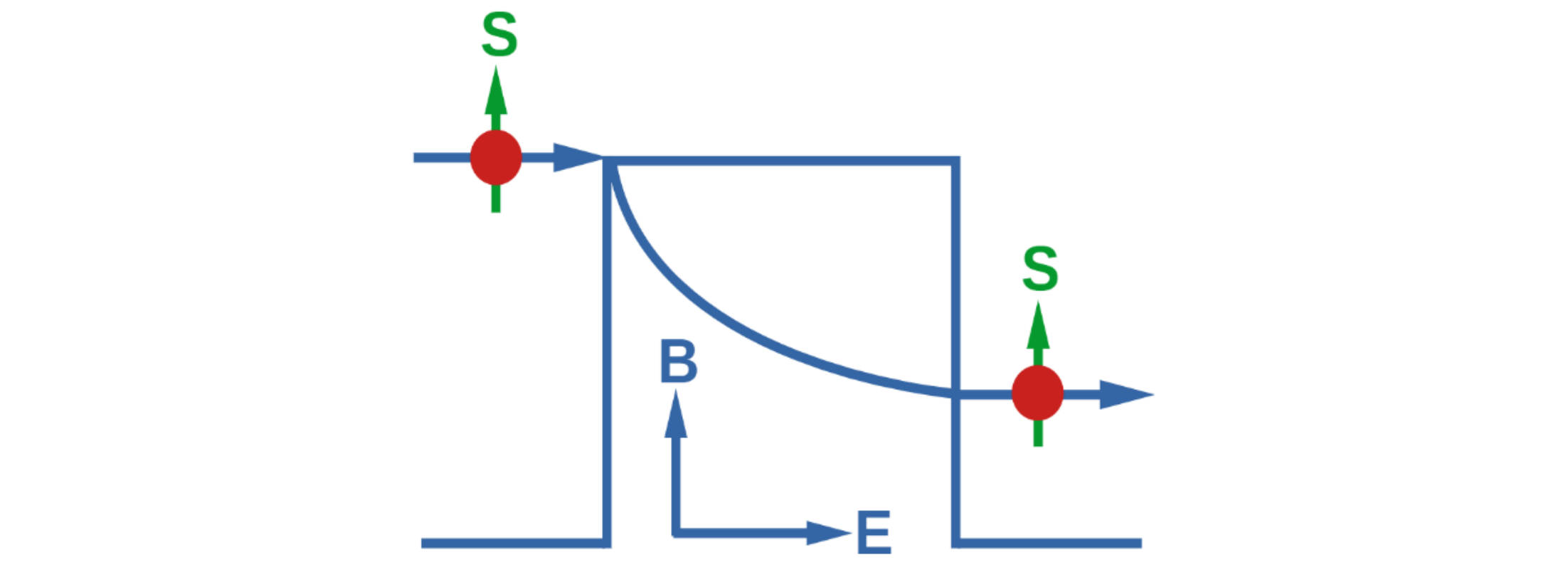

**Figure 2**: Quantum tunneling of a Dirac particle in a degenerate state.

Furthermore, according to Theorem 5.4 in [3], the spinors described by equations (3.7) and (3.8) will also be solutions to the Dirac equation for the 4-potentials

$$(b_0, b_1, b_2, b_3) = s(\mathbf{r}, t)(1, 0, 1, 0) \tag{3.9}$$

corresponding to the following electromagnetic fields

$$\mathbf{E} = -\frac{\partial s_q}{\partial x}\mathbf{i} + \left(\frac{\partial s_q}{\partial t} - \frac{\partial s_q}{\partial y}\right)\mathbf{j} - \frac{\partial s_q}{\partial z}\mathbf{k}, \qquad \mathbf{B} = \frac{\partial s_q}{\partial z}\mathbf{i} - \frac{\partial s_q}{\partial x}\mathbf{k} \tag{3.10}$$

where $s_q = s(\mathbf{r}, t)/q$.

For example, setting

$$s_q = -E_{W1}\cos\left[k_W(y+t) + \delta_{W1}\right]x - E_{W2}\left[k_W(y+t) + \delta_{W2}\right]z \tag{3.11}$$

in Eq. (3.10), the resulting electromagnetic field becomes

$$\mathbf{E}_W = E_{W1}\cos\left[k_W(y+t) + \delta_{W1}\right]\mathbf{i} + E_{W2}\cos\left[k_W(y+t) + \delta_{W2}\right]\mathbf{k} \tag{3.12}$$

$$\mathbf{B}_W = -E_{W2}\cos\left[k_W(y+t) + \delta_{W2}\right]\mathbf{i} + E_{W1}\cos\left[k_W(y+t) + \delta_{W1}\right]\mathbf{k} \tag{3.13}$$

describing a plane electromagnetic wave, of arbitrary polarization, propagating along the $y$ direction, with Poynting vector

$$\begin{aligned}\mathbf{S} &= \frac{1}{4\pi}\mathbf{E}_W \times \mathbf{B}_W \\ &= -\frac{1}{4\pi}\left[E_{W1}{}^2\cos^2\left[k_W\left(y+t\right)+\delta_{W1}\right]+E_{W2}{}^2\cos^2\left[k_W\left(y+t\right)+\delta_{W2}\right]\right]\mathbf{j}\end{aligned} \tag{3.14}$$

In the above expressions $E_{W1}, \delta_{W1}, E_{W2}, \delta_{W2}$ are real constants corresponding to the amplitude and phase of the x and z component of the electric field of the wave respectively, and $k_W$ is a real parameter corresponding to the wavenumber. It should also be mentioned that the $y$ direction can be set to correspond to any desired direction in space, perpendicular to the direction of motion of the particles. Thus, the state of the particles inside the potential barrier, and consequently the transmittance though the barrier, will not change in the presence of a plane electromagnetic wave, e.g. a laser beam, with arbitrary polarization, propagating along a direction perpendicular to the direction of motion of the particles.

Generally, the property of particles described by degenerate spinors to be in the same state under a wide variety of electromagnetic fields gives us the opportunity to manipulate the motion of the particles in free space using appropriate fields, without affecting their state inside the potential barrier, and consequently, the transmittance through the barrier. More details can be found in [37].

## 4. Degenerate wave-like solutions to the Dirac equation for massive particles

In the previous section we have discussed degenerate solutions for massive Dirac particles. However, these solutions are localized, describing particles existing only in classically forbidden regions. Therefore, an interesting question is the following: Are there non-localized degenerate solutions for massive particles, able to exist throughout space and time?

Indeed, in [38] it has been shown that all spinors of the form

$$\Psi = c_1 \exp\left(ih\right)\begin{pmatrix}\cos\alpha \\ \sin\alpha\exp\left(id\right) \\ \cos\beta \\ \sin\beta\exp\left(id\right)\end{pmatrix} \tag{4.1}$$

are degenerate solutions to the Dirac equation for the following 4-potential

$$\begin{pmatrix} a_0 \\ a_1 \\ a_2 \\ a_3 \end{pmatrix} = \begin{pmatrix} -\dfrac{\tan(\alpha+\beta)\left(\left(\sin(2\alpha)-\sin(2\beta)\right)\dfrac{\partial h}{\partial t}+m\left(\sin(2\alpha)+\sin(2\beta)\right)\right)}{\cos(2\alpha)-\cos(2\beta)} \\ -2m\cos\alpha\cos\beta\csc(\alpha-\beta)\sec(\alpha+\beta)\cos d+\dfrac{\partial h}{\partial x} \\ -2m\cos\alpha\cos\beta\csc(\alpha-\beta)\sec(\alpha+\beta)\sin d+\dfrac{\partial h}{\partial y} \\ \dfrac{\partial h}{\partial z} \end{pmatrix} \tag{4.2}$$

where $c_1$ is an arbitrary complex constant, $\alpha,\beta$ are real constants, $h$ is an arbitrary real functions of the spatial coordinates and time and

$$d = \frac{4m\left[t - z\cos(\alpha+\beta)\right]}{\cos(2\alpha)-\cos(2\beta)}. \tag{4.3}$$

In the above expressions we have also supposed that $\cos(2\alpha)-\cos(2\beta)\neq 0$, $\cos(\alpha+\beta)\neq 0$, $\sin(\alpha-\beta)\neq 0$, implying that $\alpha\pm\beta\neq n\pi$ and $\alpha+\beta\neq n\pi+\pi/2$, $n\in\mathbb{Z}$.

An important characteristic of these solutions is that they can describe particles of any mass, including massless particles. In addition, they correspond to particles in non-localized states, which can exist throughout space and time without any restriction, contrary to the degenerate solutions provided in [37], which describe particles existing only in classically forbidden regions, e.g. in quantum tunnelling.

Another interesting remark is that the electromagnetic 4-potentials and fields corresponding to these solutions, if the following conditions are valid:

$$\frac{\partial h}{\partial x} = \frac{\partial h}{\partial y} = \frac{\partial h}{\partial z} = 0 \tag{4.4}$$

$$\alpha = n\pi + \frac{\pi}{2} \text{ or } \beta = n\pi + \frac{\pi}{2},\quad n\in\mathbb{Z} \tag{4.5}$$

and

$$\frac{\partial h}{\partial t} = -m\frac{\sin(2\alpha)+\sin(2\beta)}{\sin(2\alpha)-\sin(2\beta)} \tag{4.6}$$

For example, the spinors

$$\Psi_0 = c_1\exp(imt)\begin{pmatrix} 0 \\ \exp(id) \\ \cos\beta \\ \sin\beta\exp(id) \end{pmatrix},\qquad \Psi'_0 = c_1\exp(-imt)\begin{pmatrix} \cos\alpha \\ \sin\alpha\exp(id) \\ 0 \\ \exp(id) \end{pmatrix} \tag{4.7}$$

describe particles in degenerate states that can exist in a region of space free of electromagnetic potential and fields.

Furthermore, the electromagnetic fields corresponding to the 4-potential given by Eq. (4.2) are of the form

$$\mathbf{E}=\frac{4m^2}{q}\cos\alpha\cos\beta\csc^2(\alpha-\beta)\csc(\alpha+\beta)\sec(\alpha+\beta)(-\sin d\ \mathbf{i}+\cos d\ \mathbf{j}) \quad (4.8)$$

$$\mathbf{B}=-\frac{4m^2}{q}\cos\alpha\cos\beta\csc^2(\alpha-\beta)\csc(\alpha+\beta)(\cos d\ \mathbf{i}+\sin d\ \mathbf{j}) \quad (4.9)$$

resembling a circularly polarized wave propagating along the +z-direction with Poynting vector

$$\mathbf{S}=\frac{1}{4\pi}\mathbf{E}\times\mathbf{B}=\frac{4m^4}{\pi q^2}\cos^2\alpha\cos^2\beta\csc^4(\alpha-\beta)\csc^2(\alpha+\beta)\sec(\alpha+\beta)\mathbf{k} \quad (4.10)$$

In addition, according to Theorem 5.4 in [3], the spinors described by Eq. (4.1) will also be solutions to the Dirac equation for an infinite number of 4-potentials, given by the formula

$$b_\mu = a_\mu + s\kappa_\mu \quad (4.11)$$

where

$$\begin{aligned}(\kappa_0,\kappa_1,\kappa_2,\kappa_3)&=\left(1,-\frac{\Psi^T\gamma^0\gamma^1\gamma^2\Psi}{\Psi^T\gamma^2\Psi},-\frac{\Psi^T\gamma^0\Psi}{\Psi^T\gamma^2\Psi},\frac{\Psi^T\gamma^0\gamma^2\gamma^3\Psi}{\Psi^T\gamma^2\Psi}\right)\\&=\left(1,-\sin(\alpha+\beta)\cos d,-\sin(\alpha+\beta)\sin d,-\cos(\alpha+\beta)\right)\end{aligned} \quad (4.12)$$

and $s$ is an arbitrary real function of the spatial coordinates and time.

For example, if the arbitrary function $s$ is constant, the electromagnetic fields corresponding to the 4-potential $b_\mu - a_\mu$ become

$$\begin{aligned}\mathbf{E}_s=&\frac{2m}{q}\csc(\alpha-\beta)\left(2m\cos\alpha\cos\beta\csc(\alpha-\beta)\csc(\alpha+\beta)\sec(\alpha+\beta)+s\right)\\&\times(-\sin d\ \mathbf{i}+\cos d\ \mathbf{j})\end{aligned} \quad (4.13)$$

$$\begin{aligned}\mathbf{B}_s=&-\frac{2m}{q}\csc(\alpha-\beta)\left(2m\cos\alpha\cos\beta\csc^2(\alpha-\beta)\csc(\alpha+\beta)+s\cos(\alpha+\beta)\right)\\&\times(\cos d\ \mathbf{i}+\sin d\ \mathbf{j})\end{aligned} \quad (4.14)$$

having the same spatial and temporal dependence with the electromagnetic fields given by equations (4.8) and (4.9). This practically means that the state of the particles does not depend on the magnitude of the fields, but only on their spatial and temporal dependence, defined by the parameter

$$d = \frac{4m\left[t - z\cos\left(\alpha + \beta\right)\right]}{\cos\left(2\alpha\right) - \cos\left(2\beta\right)} = \omega_d t - k_d z \tag{4.15}$$

where

$$\omega_d = \frac{4m}{\cos\left(2\alpha\right) - \cos\left(2\beta\right)} \tag{4.16}$$

and

$$k_d = \frac{4m\cos\left(\alpha + \beta\right)}{\cos\left(2\alpha\right) - \cos\left(2\beta\right)} \tag{4.17}$$

are constants, corresponding to the angular frequency and the wavenumber, respectively. This result indicates that the frequency of these wave-like fields depends on the mass of the particles and takes values of the order of $5 \times 10^{20}$ Hz in the case of electrons, corresponding to photons with energy higher than 2 MeV, in the region of Gamma/X-rays. Obviously, these values increase dramatically for heavier particles, e.g. protons [38].

Another interesting remark is that the phase velocity $\upsilon_{ph} = \omega_d / k_d = \sec\left(\alpha + \beta\right)$ corresponding to the wave-like fields given by equations (4.8) and (4.9) is higher than the speed of light ( $c = 1$ in natural units), albeit without violating the special theory of relativity, since a sinusoidal wave with a unique frequency does not transmit any information. Finally, it should be mentioned that the expected values of the projections of the spin of the particles along the x, y, and z-axes, defined as [38, 39]

$$S_x = \frac{i}{2}\Psi^\dagger \gamma^2 \gamma^3 \Psi = \frac{\left|c_1\right|^2}{2}\left(\sin\left(2\alpha\right) + \sin\left(2\beta\right)\right)\cos d \tag{4.18}$$

$$S_y = \frac{i}{2}\Psi^\dagger \gamma^3 \gamma^1 \Psi = \frac{\left|c_1\right|^2}{2}\left(\sin\left(2\alpha\right) + \sin\left(2\beta\right)\right)\sin d \tag{4.19}$$

$$S_z = \frac{i}{2}\Psi^\dagger \gamma^1 \gamma^2 \Psi = \frac{\left|c_1\right|^2}{2}\left(\cos\left(2\alpha\right) + \cos\left(2\beta\right)\right) \tag{4.20}$$

are all synchronized with the magnetic component of the electromagnetic fields corresponding to these solutions [38]. This result indicates that the synchronization between the spin of the particles and the electromagnetic fields corresponding to the degenerate spinors can be regarded as a key feature of these solutions. This is further enhanced by the fact that the rotation of the spin of the particles occurs even in the absence of electromagnetic fields [38].

## 5. A general method for obtaining degenerate solutions to the Dirac and Weyl equations

In this section we discuss some general forms of degenerate solutions to the Dirac and Weyl equations, calculating the degenerate spinors corresponding to the massive Dirac, massless Dirac and Weyl equations for a given set of real 4-potentials. In more detail, in [32] we have shown that all spinors of the form

$$\begin{aligned}\tilde{\Psi} &= \exp\left(i\int \tilde{f}_{1I}(s_0,s_1)\,ds_1 + \tilde{f}_{2R}(s_0)(s_2+s_3) + i\tilde{f}_{2I}(s_0)(s_2-s_3)\right)\\ &\times g(s_0)\exp\left(-\frac{m^2\cos^2\varphi}{k}s_2\right)\exp(ks_3)\\ &\times\left(i\frac{m(1+\sin\varphi)}{k}\begin{pmatrix}\cos\varphi\\ 1-\sin\varphi\\ \cos\varphi\\ 1-\sin\varphi\end{pmatrix}+\begin{pmatrix}-\cos\varphi\\ 1+\sin\varphi\\ \cos\varphi\\ -1-\sin\varphi\end{pmatrix}\right)\end{aligned} \tag{5.1}$$

where $m$ is the mass of the particles, $g(s_0)$ is an arbitrary complex function of $s_0$, $k$ is an arbitrary complex constant and $\varphi \neq n\pi + \pi/2,\ n\in\mathbb{Z}$ is a real constant, are degenerate solutions to the Dirac equation for the following 4-potentials:

$$\begin{aligned}a_0 &= h\\ a_1 &= -h\cos\varphi + f_{1I}\sec\varphi + f_{2I}\tan\varphi\\ a_2 &= f_{2R}\sec\varphi\\ a_3 &= -h\sin\varphi - f_{2I}\end{aligned} \tag{5.2}$$

where $f_{1I}$, $f_{2R}$, $f_{2I}$, $h$ are arbitrary real functions of the spatial coordinates and time. The functions $\tilde{f}_{1I}(s_0,s_1), \tilde{f}_{2R}(s_0), \tilde{f}_{2I}(s_0)$ are related to $f_{1I}$, $f_{2R}$, $f_{2I}$ through the following coordinate transformation [32]:

$$\begin{pmatrix}s_1\\ s_2\\ s_3\\ s_0\end{pmatrix}=\begin{pmatrix}\sec\varphi & 0 & 0 & 0\\ \frac{1}{2}\tan\varphi & \frac{i}{2}\sec\varphi & -\frac{1}{2} & 0\\ -\frac{1}{2}\tan\varphi & \frac{i}{2}\sec\varphi & \frac{1}{2} & 0\\ -\cos\varphi & 0 & -\sin\varphi & 1\end{pmatrix}\begin{pmatrix}x\\ y\\ z\\ t\end{pmatrix} \tag{5.3}$$

Thus, for any combination of the arbitrary functions $\tilde{f}_{1I}(s_0,s_1), \tilde{f}_{2R}(s_0), \tilde{f}_{2I}(s_0)$ one can automatically construct a degenerate solution to the Dirac equation corresponding to an infinite number of real 4-potentials, given by Eq. (5.2).

Some important remarks regarding these solutions are the following:

- Massive particles described by these spinors should be localized, both in space and time, because otherwise the solution would be divergent.
- All the information regarding the 4-potentials is incorporated into the phase of the spinors.
- The expected values of the projections of the spin of the particles along the x, y, and z axes are functions of the mass of the particles and the spatial and temporal coordinates.
- However, in the special case that $k = m\cos\varphi$ the expected values of the projections of the spin of the particles along the x, y, and z axes become all equal to zero.

In the case of massless particles, the degenerate solutions take the following form [32]:

$$\tilde{\Psi} = \exp\left(i\int \tilde{f}_{1I}(s_0,s_1)\,ds_1 + \tilde{f}_{2R}(s_0)(s_2+s_3) + i\tilde{f}_{2I}(s_0)(s_2-s_3)\right) \times \left( \tilde{W}_T(s_0,s_2)\begin{pmatrix} \cos\varphi \\ 1-\sin\varphi \\ \cos\varphi \\ 1-\sin\varphi \end{pmatrix} + \tilde{W}_R(s_0,s_3)\begin{pmatrix} -\cos\varphi \\ 1+\sin\varphi \\ \cos\varphi \\ -1-\sin\varphi \end{pmatrix} \right) \tag{5.4}$$

where $\tilde{W}_T(s_0,s_2)$, $\tilde{W}_R(s_0,s_3)$ are arbitrary complex functions of the coordinates $s_0,s_2$ and $s_0,s_3$ respectively. Obviously, special care must be taken to ensure that the spinors given by Eq. (5.4) is bound for all values of the spatial coordinates and time. The simplest choice to satisfy this condition is setting the functions $\tilde{W}_T(s_0,s_2)$, $\tilde{W}_R(s_0,s_3)$ to be wave like, taking the following form:

$$\tilde{W}_T(s_0,s_2) = c_T \exp\left(-ik_I\left(t - x\cos\varphi - z\sin\varphi\right)\right) \tag{5.5}$$

$$\tilde{W}_R(s_0,s_2) = c_R \exp\left(-ik_I\left(t - x\cos\varphi - z\sin\varphi\right)\right) \tag{5.6}$$

where $c_T, c_R$ are arbitrary complex constants. As an example, we consider the spinor

$$\Psi = \exp\Big(i\sec\varphi\big((k_1+k_2\sin\varphi)x + k_3 y - k_2 z\cos\varphi\big)(t - x\cos\varphi - z\sin\varphi)\Big)$$
$$\times\exp\big(-ik_I(t-x\cos\varphi - z\sin\varphi)\big)\left(c_T\begin{pmatrix}\cos\varphi\\ 1-\sin\varphi\\ \cos\varphi\\ 1-\sin\varphi\end{pmatrix} + c_R\begin{pmatrix}-\cos\varphi\\ 1+\sin\varphi\\ \cos\varphi\\ -1-\sin\varphi\end{pmatrix}\right) \quad (5.7)$$

which is a degenerate solution to the massless Dirac equation for the set of real 4-potentials given by Eq. (5.2).

Here, an interesting remark is that, unlike the case of massive particles, Dirac spinors corresponding to massless particles do not require localization. Consequently, massless Dirac particles in degenerate states can move freely through space and time without any restrictions. On the other hand, the information regarding the 4-potentials is encoded in the phase of the spinors, as in the case of massive particles. In addition, it should be mentioned that, in the case of massless particles, the spin projections along the x, y, and z axes remain constant.

Finally, in the case that $\tilde{W}_R(s_0,s_3)=0$ or $\tilde{W}_T(s_0,s_2)=0$, the degenerate spinors given by Eq. (5.4) take the simpler form $\tilde{\Psi}=(\tilde{\psi}_T,\tilde{\psi}_T)^T$ or $\tilde{\Psi}=(\tilde{\psi}_R,-\tilde{\psi}_R)^T$ respectively, where

$$\tilde{\psi}_T = \exp\Big(i\int \tilde{f}_{1I}(s_0,s_1)\,ds_1 + \tilde{f}_{2R}(s_0)(s_2+s_3) + i\tilde{f}_{2I}(s_0)(s_2-s_3)\Big)$$
$$\times\tilde{W}_T(s_0,s_2)\begin{pmatrix}\cos\varphi\\ 1-\sin\varphi\end{pmatrix} \quad (5.8)$$

and

$$\tilde{\psi}_R = \exp\Big(i\int \tilde{f}_{1I}(s_0,s_1)\,ds_1 + \tilde{f}_{2R}(s_0)(s_2+s_3) + i\tilde{f}_{2I}(s_0)(s_2-s_3)\Big)$$
$$\times\tilde{W}_R(s_0,s_3)\begin{pmatrix}-\cos\varphi\\ 1+\sin\varphi\end{pmatrix} \quad (5.9)$$

are solutions to the Weyl equation in the form (1.9) and (1.11) respectively [32]. As in the case of massless Dirac particles, Weyl particles can move freely throughout space and time. Additionally, it should be noted that the phase factor

$$\exp\Big(i\int \tilde{f}_{1I}(s_0,s_1)\,ds_1 + \tilde{f}_{2R}(s_0)(s_2+s_3) + i\tilde{f}_{2I}(s_0)(s_2-s_3)\Big) \quad (5.10)$$

containing the information regarding the electromagnetic 4-potentials is the same in all cases, regarding both massive and massless particles.

## 6. On the localization of Weyl particles using simple electric fields

In this section we discuss a special class of solutions to the Weyl equations having the remarkable property to describe particles in localized states, even in the absence of electromagnetic fields. In more detail, in [34], [40] it has been shown that all spinors of the form:

$$\psi = \begin{pmatrix} \cos\left(\dfrac{\theta(t)}{2}\right) \\ e^{i\varphi(t)}\sin\left(\dfrac{\theta(t)}{2}\right) \end{pmatrix} \exp\left[ih(\mathbf{r},t)\right] \tag{6.1}$$

are solutions to the Weyl equation (1.9) corresponding to particles with positive helicity for the following 4-potential:

$$(a_0,a_1,a_2,a_3) = \left(\frac{\partial h}{\partial t}+\frac{1}{2}\frac{d\varphi}{dt}, \frac{\partial h}{\partial x}+\frac{1}{2}\sin\varphi\frac{d\theta}{dt}, \frac{\partial h}{\partial y}-\frac{1}{2}\cos\varphi\frac{d\theta}{dt}, \frac{\partial h}{\partial z}-\frac{1}{2}\frac{d\varphi}{dt}\right) \tag{6.2}$$

where $\theta(t),\varphi(t)$ are arbitrary real functions of time and $h(\mathbf{r},t)$ is any real function of the spatial coordinates and time. Similar expressions can be obtained for particles with negative helicity [40].

The electromagnetic field corresponding to the 4-potential $a_\mu$ is given by the following formula:

$$\begin{aligned} \mathbf{E} &= \frac{1}{2q}\left(\cos\varphi\frac{d\theta}{dt}\frac{d\varphi}{dt}+\sin\varphi\frac{d^2\theta}{dt^2}\right)\mathbf{i}+\frac{1}{2q}\left(\sin\varphi\frac{d\theta}{dt}\frac{d\varphi}{dt}-\cos\varphi\frac{d^2\theta}{dt^2}\right)\mathbf{j}-\frac{1}{2q}\frac{d^2\varphi}{dt^2}\mathbf{k} \\ \mathbf{B} &= \mathbf{0} \end{aligned} \tag{6.3}$$

Consequently, in the case that

$$\frac{d^2\theta}{dt^2}=\frac{d^2\varphi}{dt^2}=\frac{d\theta}{dt}\frac{d\varphi}{dt}=0 \tag{6.4}$$

$\mathbf{E}=\mathbf{B}=\mathbf{0}$, implying that Weyl particles in the absence of electromagnetic fields can exhibit one of the following behaviors:

- move as free particles, assuming that $(d\theta/dt = d\varphi/dt = 0)$
- exist in a localized bounded state assuming that $(d\theta/dt=\omega_1,\ d\varphi/dt=0)$
- exist in an intermediate state, bound on the x-y plane, and moving freely along the z-axis, in the case that $(d\theta/dt=0,\ d\varphi/dt=\omega_2)$

Furthermore, the localization of the particles can be easily controlled using simple constant electric fields [40], as shown in the following figures.

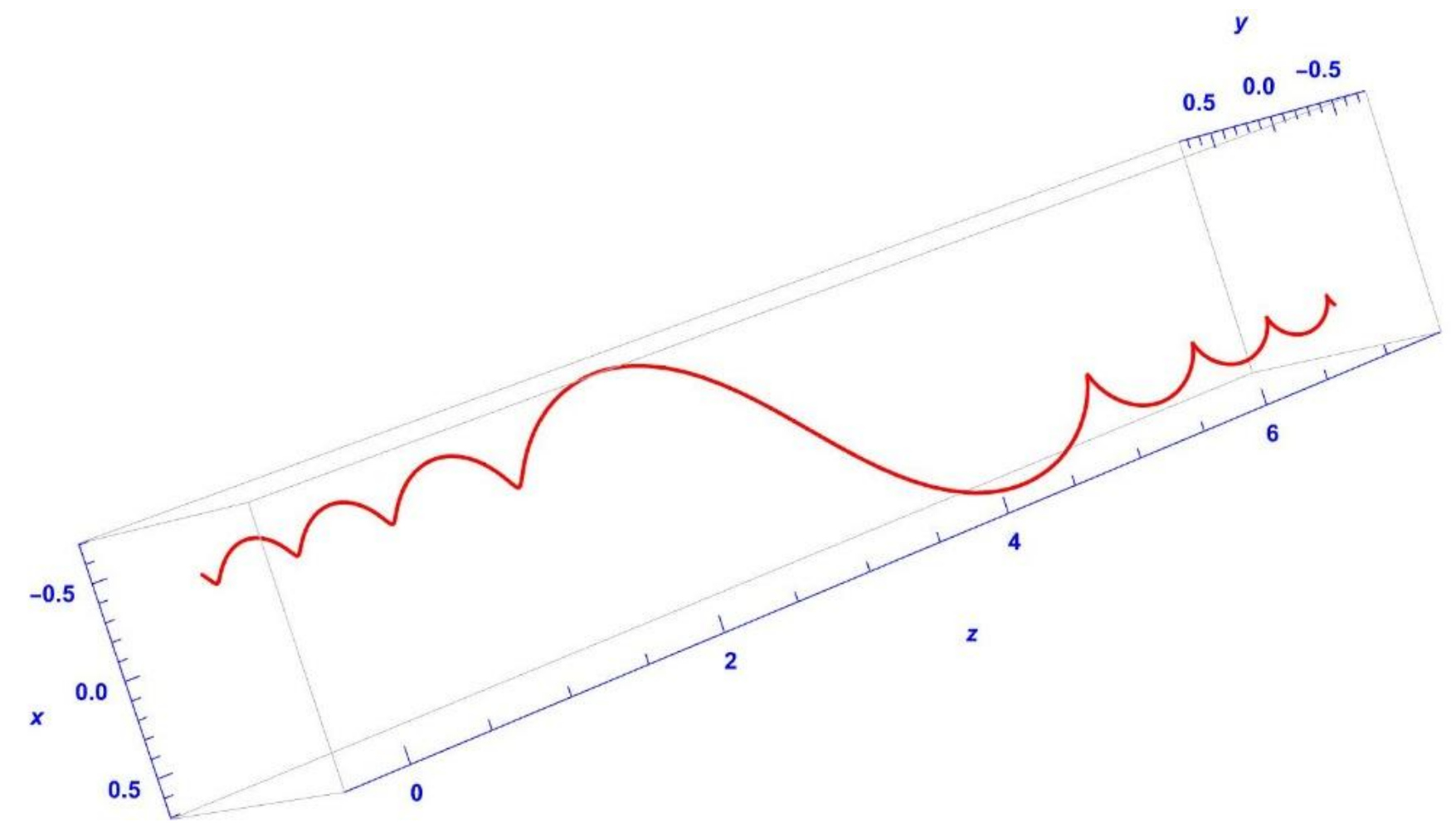

**Figure 3**: The trajectory of a classical particle with the same velocity as the Weyl particle for $\theta(t)=\pi/4$ and $\varphi(t)=10t-t^2$, corresponding to a constant electric field $\mathbf{E}=(1/q)\mathbf{k}$, applied for $t\in[0,10]$ [40].

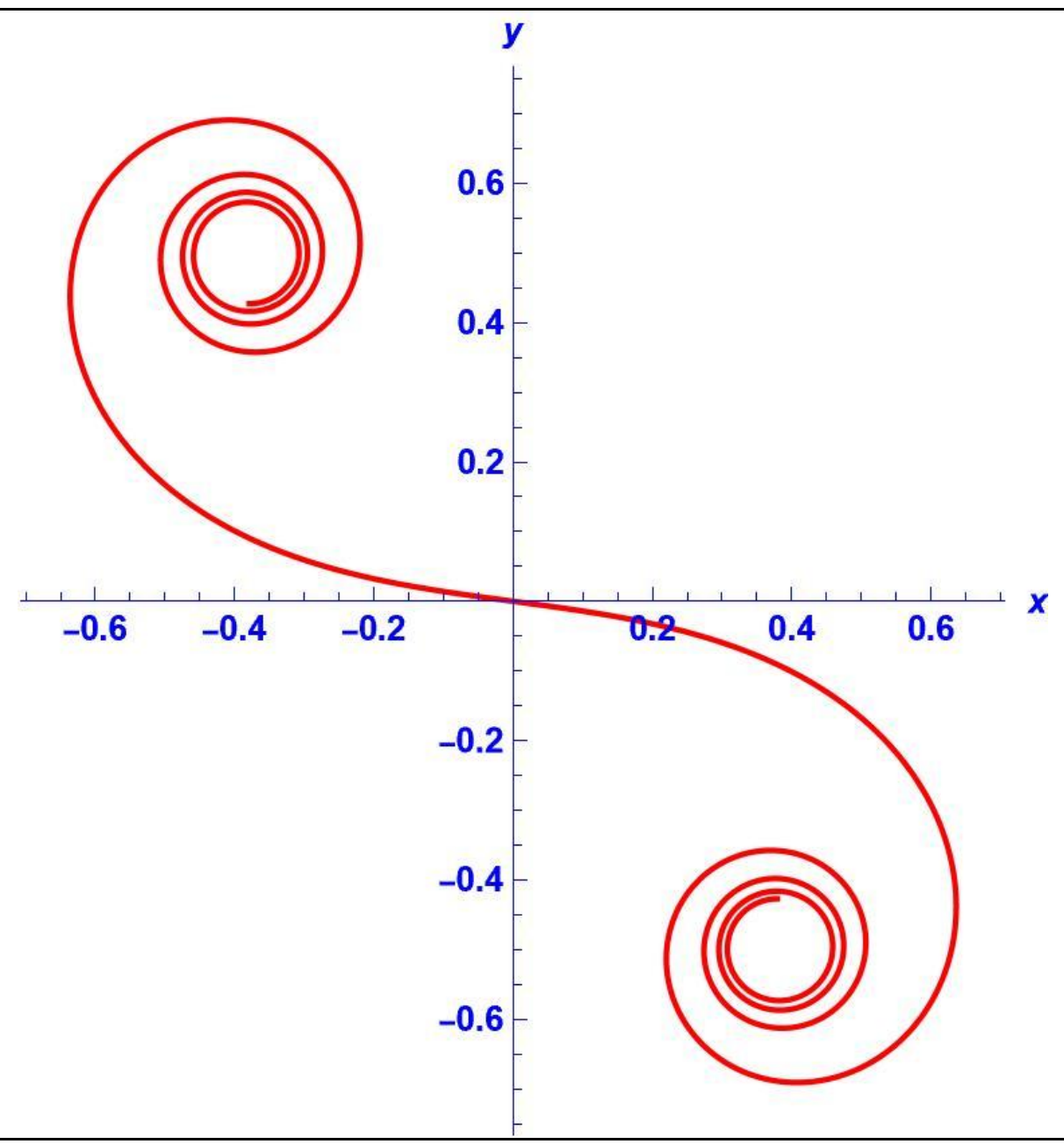

**Figure 4**: The projection of the motion of a classical particle with the same velocity as the Weyl one, on the x-y plane. The parameters used in this simulation are the same as in the previous figure [40].

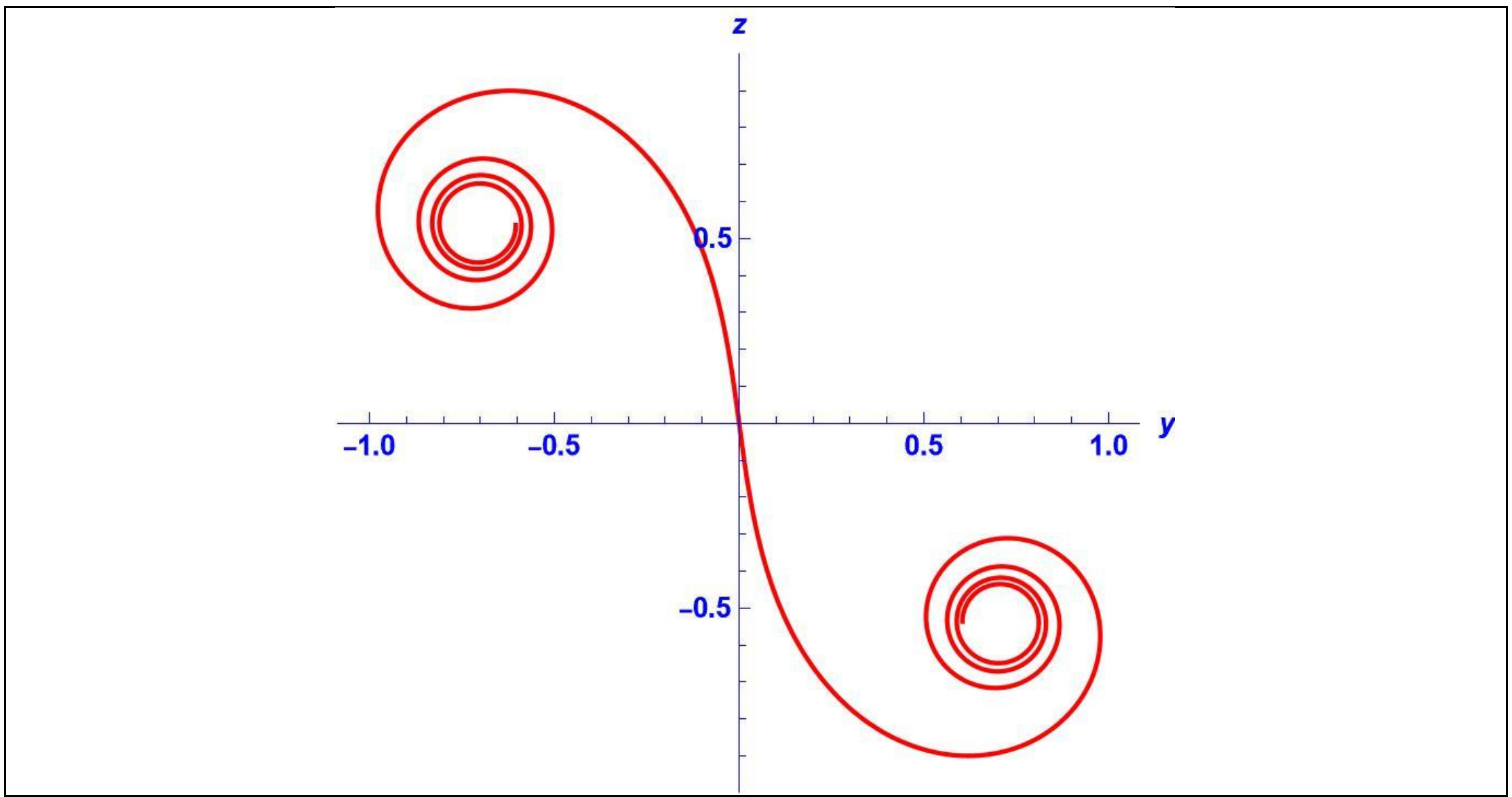

**Figure 5**: The trajectory of a classical particle with the same velocity as the Weyl one for $\varphi(t)=\pi/2$, $\theta(t)=10t-t^2$ and $t\in[0,10]$, corresponding to a constant electric field $\mathbf{E}=(1/q)\mathbf{i}$. The motion of the particle is restricted on the $y$-$z$ plane, perpendicular to the applied electric field [40].

Here, it should also be noted that the time required for the localization of the particles in a region of radius $r_0$ is of the order of

$$\Delta t=\frac{\hbar}{2qr_0|\mathbf{E}|} \tag{6.5}$$

in S.I. units [40], where $|\mathbf{E}|$ is the magnitude of the electric field. It has also been assumed that the particle is fully delocalized at $t=0$ $\left(r_0(0)\to\infty\right)$.

## 7. A proposed device for controlling the flow of information based on Weyl Fermions

Based on these results we have proposed a novel device for controlling the flow of information at a rate of up to 100 Petabits per second using Weyl Fermions [36]. The proposed device consists of a slab of material supporting Weyl particles. An array of capacitors, each corresponding to an information channel, is constructed on this material to control the motion of Weyl fermions in each channel. If no voltage is applied to the capacitors, Weyl particles move along straight lines on each channel, generating a logical “one” in the channel output. On the other hand, if a voltage is applied to selected channels of the capacitor, the resulting electric field disrupts the motion of the particles in these channels generating a logical “zero”. Consequently, we can easily control the flow of information through the channels, as shown in figure 6.

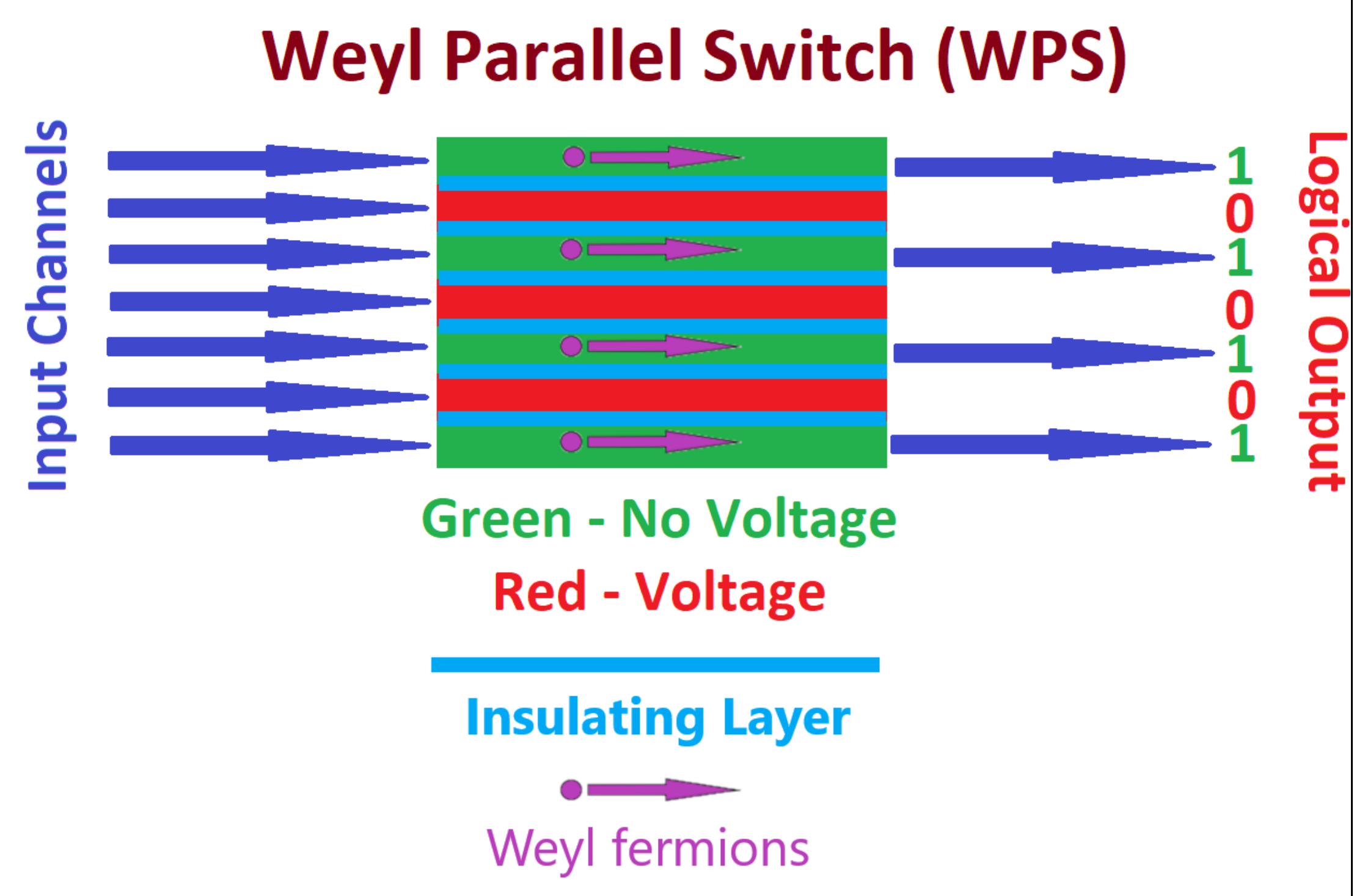

**Figure 6**: Schematic diagram of a device for controlling the flow of information based on Weyl particles [36].

This approach enables rapid switching between logical "zeros" and "ones" with a response time on the order of 1 ps for typical values of the parameters, as given by Eq. (6.5) [36]. Moreover, if each channel has a width of the order of 100 nm and the total width of the material used in this system is of the order of 1 cm, the device can accommodate up to $10^5$ channels. As a result, it is possible to control the flow of information at a rate of the order of 100 petabits per second, a level of performance that is extremely challenging to achieve with conventional electronics.

Additionally, replacing electrons with Weyl particles for information transfer provides significant additional advantages, including much higher transmission speeds—twice as fast as in graphene and up to 1000 times greater than those found in traditional semiconductors [10, 20]. This also leads to improved energy efficiency, as Weyl particles experience fewer collisions with lattice ions, thereby minimizing heat generation. Consequently, the power consumption of this device, as well as any other device utilizing Weyl particles, is expected to be several orders of magnitude lower than that of standard electronic devices.

In addition, as shown in [3], Weyl particles have the remarkable property to be able to exist in the same quantum state, under a wide variety of electromagnetic fields. Thus, the proposed device is expected to offer enhanced robustness against electromagnetic perturbations.

It should also be mentioned that WPS could operate as an electric field sensor. In more detail, the presence of an electric field perpendicular to the Weyl current propagating in a specific channel of the device, could alter the propagation direction of Weyl particles, interrupting the current in this channel. Thus, it is possible to detect the presence of electric fields with exceptionally high spatial and temporal resolution.

Finally, in [36] we have also shown that we can fully control the transverse spatial distribution $f\left(x,y\right)$ of Weyl particles using the following magnetic field

$$\mathbf{B}=-\frac{1}{q}\frac{1}{f^2}\left[\left(\frac{\partial f}{\partial x}\right)^2+\left(\frac{\partial f}{\partial y}\right)^2-f\left(\frac{\partial^2 f}{\partial x^2}+\frac{\partial^2 f}{\partial y^2}\right)\right]\mathbf{k} \tag{7.1}$$

applied along with the direction of motion of the particles. This result provides the opportunity to guide Weyl fermions through the proposed device, or any other device utilizing Weyl particles.

## 8. Conclusions

In this work, we have explored key findings regarding the electromagnetic interactions of Dirac and Weyl particles and specifically the presence of degenerate states, where the particles can exist in the same quantum state across a broad range of electromagnetic 4-potentials and fields, which are explicitly calculated. Our results demonstrate that all solutions to the Weyl equations are degenerate, implicating that Weyl particles can exist in the same quantum state in a wide variety of electromagnetic 4-potentials and fields. We have also described degenerate solutions to the Dirac equation, especially in the framework of massless particles and quantum tunnelling.

Additionally, we have shown that Weyl particles can form localized states even in the absence of electromagnetic fields and that their localization can be finely controlled through simple electric fields. This tunability provides a promising foundation for practical applications in quantum information processing and electronic devices. Building on these findings, we have proposed an innovative device capable of controlling the flow of information at rates reaching 100 petabits per second using Weyl fermions, surpassing the limits of conventional semiconductor technology.

We have also introduced a general method for obtaining degenerate solutions, applicable universally to Dirac and Weyl particles, regardless of mass. We believe that these insights pave the way for future advancements in high-speed electronics, quantum computing, and novel materials supporting massless quasiparticles.